\documentclass[prx,10pt,twocolumn,reprint,showkeys,superscriptaddress,nofootbib,longbibliography,bibnotes]{revtex4-1}

\usepackage{amsmath}    
\usepackage{graphicx}   
\usepackage{verbatim}   
\usepackage{color}      
\usepackage{subfigure}  
\usepackage{hyperref}   
\usepackage{SIunits}    

\usepackage{amsmath}
\usepackage{bm}
\usepackage{bbold}
\usepackage{float}
\usepackage{xfrac}

\newcommand{\alphaIN}{\lvert\alpha_{\text{in}}\rvert}

\newcommand{\etaQPA}{\eta_{\text{QPA}}}
\newcommand{\etaR}{\eta_{\text{rest}}}
\newcommand{\etaMeas}{\eta_{\text{meas}}}

\newcommand{\GammaMeas}{\Gamma_\text{meas}}

\newcommand{\GammaPhi}{\Gamma_{\phi}}

\newcommand{\GammaPhiQL}{\Gamma_{\phi,\text{QL}}}
\newcommand{\GammaPhiPara}{\Gamma_{\phi,\text{parasitic}}}

\newcommand{\omegaQPA}{\omega_\text{QPA}}

\newcommand{\GQPA}{G_\text{QPA}}

\newcommand{\sigmaZ}{\hat{\sigma}_z}

\newcommand{\omegaQ}{\omega_\text{q}}
\newcommand{\SNR}{\text{SNR}}

\newcommand{\FigOne}{Fig.~\ref{intro_fig}}
\newcommand{\FigTwo}{Fig.~\ref{parasiteFig}}
\newcommand{\FigThree}{Fig.~\ref{dephaseFig}}
\newcommand{\FigFour}{Fig.~\ref{etaFig}}

\newcommand{\abs}[1]{\left|#1\right|}

\newcommand{\bra}[1]{\langle{#1}|}
\newcommand{\ket}[1]{|{#1}\rangle}

\newcommand{\QNLaffil}{\affiliation{Quantum Nanoelectronics Laboratory, Department of Physics, University of California, Berkeley CA 94720, USA.}}
\newcommand{\CQCSaffil}{\affiliation{Center for Quantum Coherent Science, University of California, Berkeley CA 94720, USA.}}

\newcommand{\mcgillAffil}{\affiliation{Department of Physics, McGill University, Montreal, Quebec H3A 2T8, Canada}}
\newcommand{\IMEaffil}{\affiliation{Institute for Molecular Engineering, University of Chicago, Chicago, Illinois 60637, USA}}

\newcommand{\QNLauthor}[1]{\author{#1}\QNLaffil\CQCSaffil}

\begin{document}

\title{High-efficiency measurement of an artificial atom embedded in a parametric amplifier}

\author{A. Eddins}
\altaffiliation{Author to whom correspondence should be addressed: aeddins@berkeley.edu}
\QNLaffil
\CQCSaffil
\QNLauthor{J.M. Kreikebaum}
\QNLauthor{D.M. Toyli}
\author{E.M. Levenson-Falk}
\altaffiliation[Current address: ]{Department of Physics \& Astronomy, University of Southern California, Los Angeles CA 90089, USA}
\QNLaffil
\CQCSaffil
\QNLauthor{A. Dove}
\QNLauthor{W.P. Livingston}
\author{B.A. Levitan}
\mcgillAffil
\author{L.C.G. Govia}
\altaffiliation[Current address: ]{Raytheon BBN Technologies, 10 Moulton St., Cambridge, MA 02138, USA}
\IMEaffil
\author{A.A. Clerk}
\IMEaffil
\QNLauthor{I. Siddiqi}

\date{\today}
\begin{abstract}

A crucial limit to measurement efficiencies of superconducting circuits comes from losses involved when coupling to an external quantum amplifier. Here, we realize a device circumventing this problem by directly embedding a two-level artificial atom, comprised of a transmon qubit, within a flux-pumped Josephson parametric amplifier. Surprisingly, this configuration is able to enhance dispersive measurement without exposing the qubit to appreciable excess backaction. This is accomplished by engineering the circuit to permit high-power operation that reduces information loss to unmonitored channels associated with the amplification and squeezing of quantum noise. By mitigating the effects of off-chip losses downstream, the on-chip gain of this device produces end-to-end measurement efficiencies of up to 80\%. Our theoretical model accurately describes the observed interplay of gain and measurement backaction, and delineates the parameter space for future improvement. The device is compatible with standard fabrication and measurement techniques, and thus provides a route for definitive investigations of fundamental quantum effects and quantum control protocols.

\end{abstract}
\maketitle
\SIunits[thinspace,thinqspace]

\section{Introduction}
The sum of interactions between a quantum system and all environmental channels produces a continuous flow of quantum information into the environment, causing dephasing at a rate $\GammaPhi$. For a two-level qubit described by $\sigmaZ$ and measured along that axis, one may define the fraction of this information flux experimentally captured per unit time to be the measurement efficiency $\etaMeas = \GammaMeas/2\GammaPhi$, a critical parameter for continuous quantum measurements, where $\GammaMeas$ represents the rate at which the experimentalist learns about $\sigmaZ$ and is defined such that $\etaMeas$ ranges from 0 to 1.

Maximizing this efficiency for superconducting qubit measurements requires multiple stages of cyrogenic amplification to boost information-bearing quantum microwaves above the noise floor of room-temperature electronics. The use of off-chip superconducting parametric amplifiers for the first gain stage has enabled a variety of experiments investigating quantum measurement dynamics \cite{Vijay2012StabilizingFeedback,Murch2013ObservingBit,Weber2014MappingStates,Hacohen-Gourgy2016QuantumObservables,Campagne-Ibarcq2016ObservingFluorescence,Ficheux2017DynamicsDephasing}, with improvements in efficiency reported using multi-junction circuits \cite{Walter2017RapidQubits}. However, prior to amplification the information encoded in the field is extremely fragile, such that in these configurations $\sim$30\% of the information is dissipated in lossy microwave circulators and other components en route to the amplifier, lowering the ceiling on $\etaMeas$. Interest in surmounting this limitation has helped spur recent progress in the development of superconducting circulators and directional amplifiers \cite{Chapman2017WidelyCircuits,Lecocq2017NonreciprocalAmplifier,Peterson2017DemonstrationCircuit,Sliwa2015ReconfigurableAmplifier,Kerckhoff2015On-ChipRotation,Ranzani2017WidebandLine,Metelmann2015NonreciprocalEngineering}. Alternatively, high efficiency may be realized by strongly measuring a second, ancillary qubit and resonator mode (e.g. \cite{Minev2018ToMid-flight}).

Here we develop a minimal circuit architecture providing on-chip parametric gain through integration of a standard Josephson parametric amplifier (JPA) with the qubit in a configuration we dub the Qubit Parametric Amplifier (QPA), removing virtually all pre-amplification loss. Previous demonstrations with on-chip amplifiers have leveraged the bifurcation dynamics of a nonlinear resonator \cite{Siddiqi2004RF-DrivenMeasurement,Schmitt2014MultiplexedAmplifiers,Krantz2016Single-shotOscillator}. In contrast, the QPA implements on-chip the parametric mode of operation that has been widely applied in continuous measurements of qubits. This scheme presents a novel challenge, as the in-situ microwave amplification and squeezing opens a parasitic measurement channel inducing excess dephasing. We model and characterize this backaction in detail, and successfully mitigate it via a weakly nonlinear design permitting fast measurement, producing steady-state efficiencies as high as $\etaMeas = 0.80$ with direction for further improvement.

\begin{figure}
\includegraphics[width=\columnwidth]{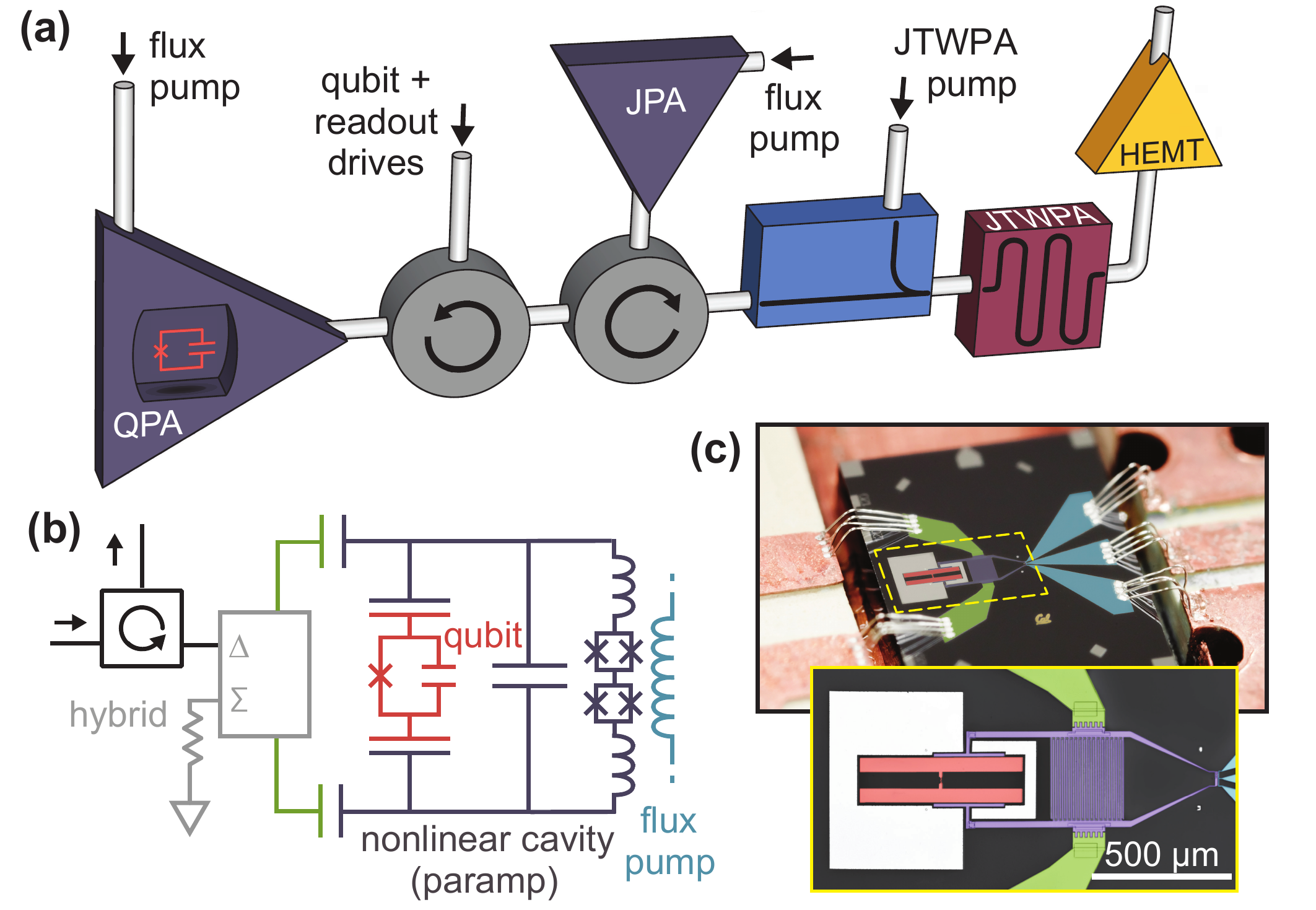}\\
\caption{\label{intro_fig} (a) Simplified experimental setup. The QPA consists of a transmon qubit dispersively coupled to a JPA acting as the readout resonator. A coherent measurement tone reflects off the QPA, carrying qubit-state information to a second, off-chip JPA followed by a Josephson Traveling Wave Parametric Amplifier (JTWPA). (b,c) Schematic and false-colored images of the QPA. The port at right (cyan) flux-couples a pump tone to the JPA, producing on-chip amplification.}
\end{figure}

A schematic of our experiment appears in \FigOne(a). The QPA consists of a transmon qubit \cite{Koch2007Charge-insensitiveBox} dispersively coupled to a JPA. A microwave readout tone at frequency $\omegaQPA$ reflects off the QPA, acquiring qubit-state information. A pump tone of the form $\cos(2(\omegaQPA t+\Phi))$ applied to the pump port of the QPA concurrent with the readout modulates the QPA resonance frequency, producing on-chip phase-sensitive amplification of the measurement field. Adjusting the phase of the pump tone relative to the readout tone changes which field quadrature is amplified and which is squeezed. The output of the QPA is then routed by microwave circulators to additional amplification stages including a second, off-chip JPA and a superconducting Josephson Traveling Wave Parametric Amplifier (JTWPA) \cite{Macklin2015AAmplifier} en route to room-temperature demodulation and digitization. By acting as a phase-sensitive preamplifier before the JTWPA, which necessarily adds at least half a photon of noise in standard phase-preserving operation, the off-chip JPA reduces the amount of on-chip gain required for high efficiency.

A circuit diagram and false-color photographs of the QPA are shown in \FigOne(b,c). The on-chip JPA design is similar to that of some off-chip JPAs \cite{Toyli2016ResonanceVacuum,Zhou2014High-gainArray}, consisting of an interdigitated capacitor in parallel with a combination of geometric and Josephson inductance to form an $LC$ resonator (purple) whose frequency $\omegaQPA$ tunes with the flux applied through the pair of SQUID loops. A superconducting coil housed below the chip enables static tuning of $\omegaQPA$, while a pump applied via the flux-line (cyan) modulates $\omegaQPA$ to produce parametric gain. Some variation in circuit parameters occurred as data were acquired over the course of multiple cooldowns; we give representative parameter values here, and list precise values for each dataset in Appendix \ref{parametersAppendix}. The QPA resonator has a zero-flux frequency of $\omega_\text{QPA,max}/2\pi = 6.970$ GHz; we tuned this down to $\omegaQPA/2\pi \leq 6.740$ GHz to increase the modulation amplitude produced by the flux-pump. Coupling capacitors and a $180\degree$ microwave hybrid couple the resonator to the readout transmission line with an effective $\kappa_\text{ext}/2\pi = 25.7$ MHz $ \gg \kappa_\text{int}/2\pi$. The transmon qubit (red) resonates at $\omegaQ/2\pi = 4.271$ GHz and is capacitively coupled to the on-chip JPA with dispersive interaction strength $\chi/2\pi = 1.9$ MHz, with the convention that the AC Stark shift changes $\omegaQ$ by $2\chi\bar{n}$. The paddle design of the qubit is chosen to reduce loss due to electromagnetic participation of the surface-vacuum interface, and a floating radiation shield (white) suppresses radiative decay of the qubit into other environmental modes. The measured lifetime $T_1 = 4.2(8)\ \mu\text{s}$ is near the expected Purcell-decay limited value $T_1 \approx 6\ \mu\text{s}$, which could be improved in future designs via integration of a Purcell filter \cite{Reed2010FastQubit}.

In the dispersive approximation and in the frame rotating at $\omegaQPA$, the internal QPA dynamics can be described by the Hamiltonian
\begin{equation}\label{eq:QPAHam}
\hat{H}_\text{QPA} \approx \frac{\hbar}{2}(\Delta+2\chi(\hat{a}^\dagger\hat{a}+1/2))\sigmaZ + \frac{i\lambda}{2}(\hat{a}^{\dagger 2}-\hat{a}^2),
\end{equation}
with $\Delta = \omegaQ - \omegaQPA$. The first of the two terms is the familiar dispersive Hamiltonian that also describes the more common case of readout using a linear resonator. The second term describes the on-chip, phase-sensitive gain process, where $\lambda$ is set by the flux-pump strength and would equal the rate of squeezing if there were no dissipation ($\kappa = 0$). A succinct theoretical analysis of the system is given in Appendix \ref{theory}, with further details available in \cite{LevitanThesis}.

\begin{figure}
  \includegraphics[width=\columnwidth]{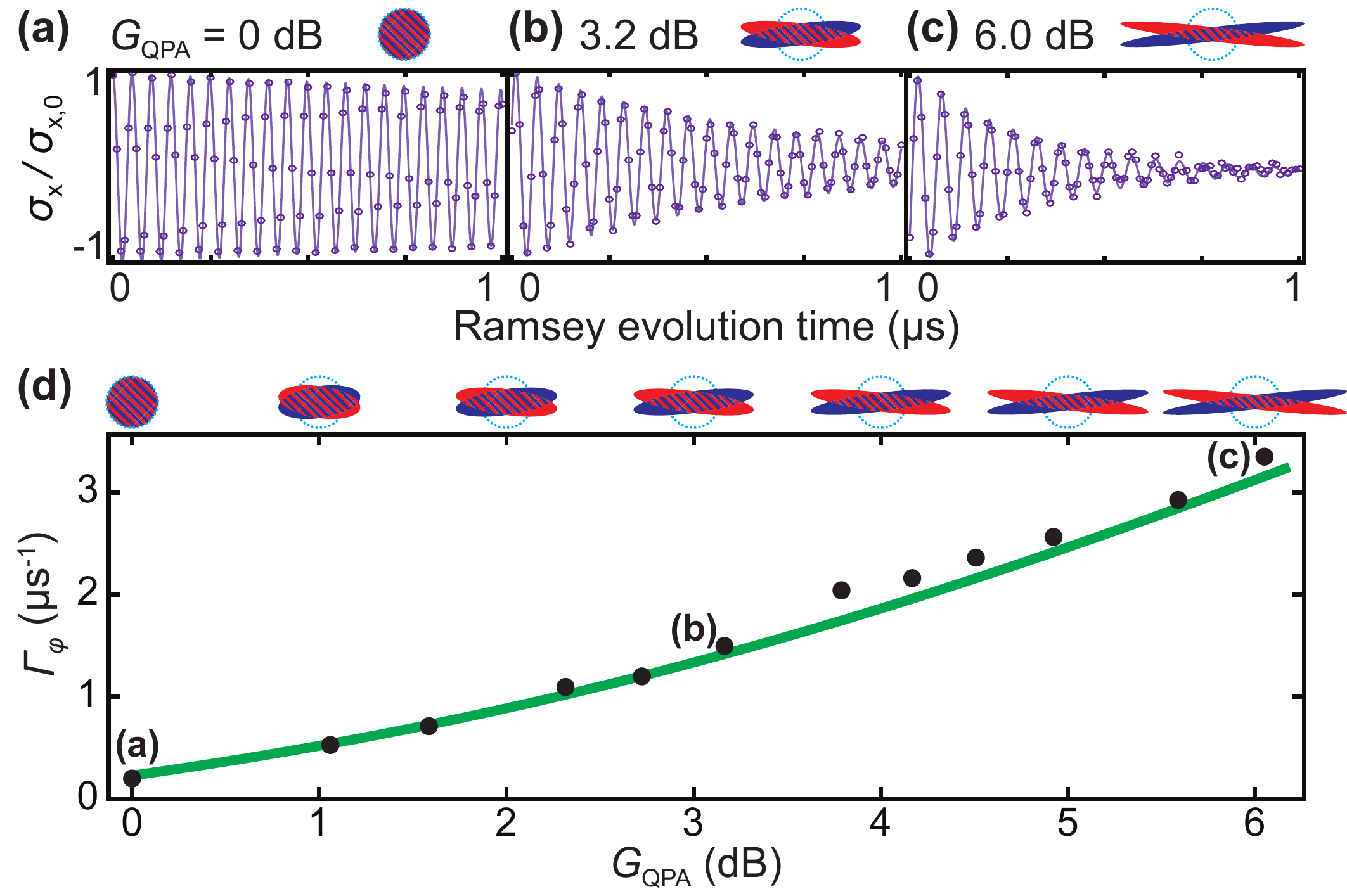}\\
  \caption{\label{parasiteFig} (a-c) Ramsey traces acquired with three values of the on-chip gain, $\GQPA$. Increasing $\GQPA$ increases the QPA output field squeezing, thus decreasing the phase-space overlap of the output fields conditioned on the ground or excited qubit states as approximately represented by the red and blue ellipses. The decreased overlap implies faster parasitic dephasing. (d) The observed dependence of $\GammaPhi$ on $\GQPA$ (black dots) is in good agreement with Eq. \ref{eq:parasiticEq} (green curve); the thickness of the curve is the standard deviation of the background dephasing rate $1/T_2^*$.}
\end{figure}

\section{Measurement backaction with on-chip gain}\label{sectDephasing}
As in conventional qubit measurement setups, the dispersive interaction encodes information about the $\sigmaZ$ component of the qubit state into the mean value of one quadrature of the output field, which we refer to as the signal quadrature $Q$. During this measurement the qubit is dephased at a rate $\GammaPhi \geq \GammaPhiQL$, where $\GammaPhiQL$ represents quantum-limited backaction. High $\etaMeas$ requires this inequality to be nearly saturated. Here, however, on-chip amplification drives a second, parasitic measurement process in which $\sigmaZ$ information is encoded in other statistical moments of the output field. This dephasing mechanism is predicted to be independent of the mean field in the resonator, making it distinct from effects in resonantly current-pumped systems \cite{Ong2011CircuitDephasing,Boissonneault2012Back-actionQubit,Ong2013QuantumQubit, Boissonneault2014SuperconductingResonator,Hatridge2011DispersiveAmplifier,Levenson-Falk2013ADetection}. A rough heuristic model describes the parasitic measurement in two steps: the phase-sensitive on-chip gain squeezes the microwave vacuum noise, and the resultant output squeezed state is rotated in phase by the dispersive interaction, encoding $\sigmaZ$ information in the covariance of the output-field quadratures. These moments are largely not detected downstream, in part due to the fragility of the moments with respect to losses, and in part because the phase-sensitive following JPA typically deamplifies this information. As the parasitic measurement increases $\GammaPhi$ without increasing the room-temperature SNR, it lowers $\etaMeas$.

Starting from Eq. \ref{eq:QPAHam}, one can derive an expression for the parasitic dephasing rate (Appendix \ref{theory}, \cite{LevitanThesis}), 
\begin{equation}\label{eq:parasiticEq}
\Gamma_{\phi,\text{parasitic}} = \frac{1}{2}\operatorname{Re}\left(\sqrt{D(-\lambda)}+\sqrt{D(\lambda)} \right)-\frac{\kappa}{2} + 1/T_2^*.
\end{equation}
Here $\lambda$ is related to the on-chip gain by
\begin{equation}
\lambda = \frac{\kappa}{2}\frac{\sqrt{\GQPA-1}}{\sqrt{\GQPA}+1},
\end{equation}
we have defined $D(\lambda)=(\kappa/2+\lambda+i\chi)^2-2i\chi\lambda$, and $T_2^*$ is an empirical parameter describing dephasing absent any applied drives. Several metrics are available to parameterize the gain dynamics; $G_\text{QPA}$ indicates the (phase-preserving) power gain experienced by a tone slightly detuned from $\omega_\text{QPA}$, which we measure directly using a vector network analyzer.

We characterize $\Gamma_{\phi,\text{parasitic}}$ via Ramsey oscillations of the qubit simultaneous with on-chip gain for several values of $G_\text{QPA}$. Absent any gain (\FigTwo(a)), we observe $1/T_2^* = 0.23(7) \mu\text{s}^{-1}$. Applying pump power produces squeezed vacuum inside the QPA, causing $\GammaPhi$ to increase significantly (\FigTwo(b,c)). The inset ellipses of \FigTwo \ indicate the quadrature variances and covariance (kurtosis is not shown) of the QPA output field predicted using equations from \cite{LevitanThesis} and experimental parameters, colored red or blue depending on the qubit state. Increasing $G_\text{QPA}$ decreases the overlap of the ellipses, speeding up the parasitic measurement. We find good agreement with the predictions of our zero-free parameter model over a range of $G_\text{QPA}$ values as plotted in \FigTwo(d), supporting the validity of the model and indicating the absence of any comparable additional dephasing mechanism for these operating conditions.

\begin{figure}
  \includegraphics[width=\columnwidth]{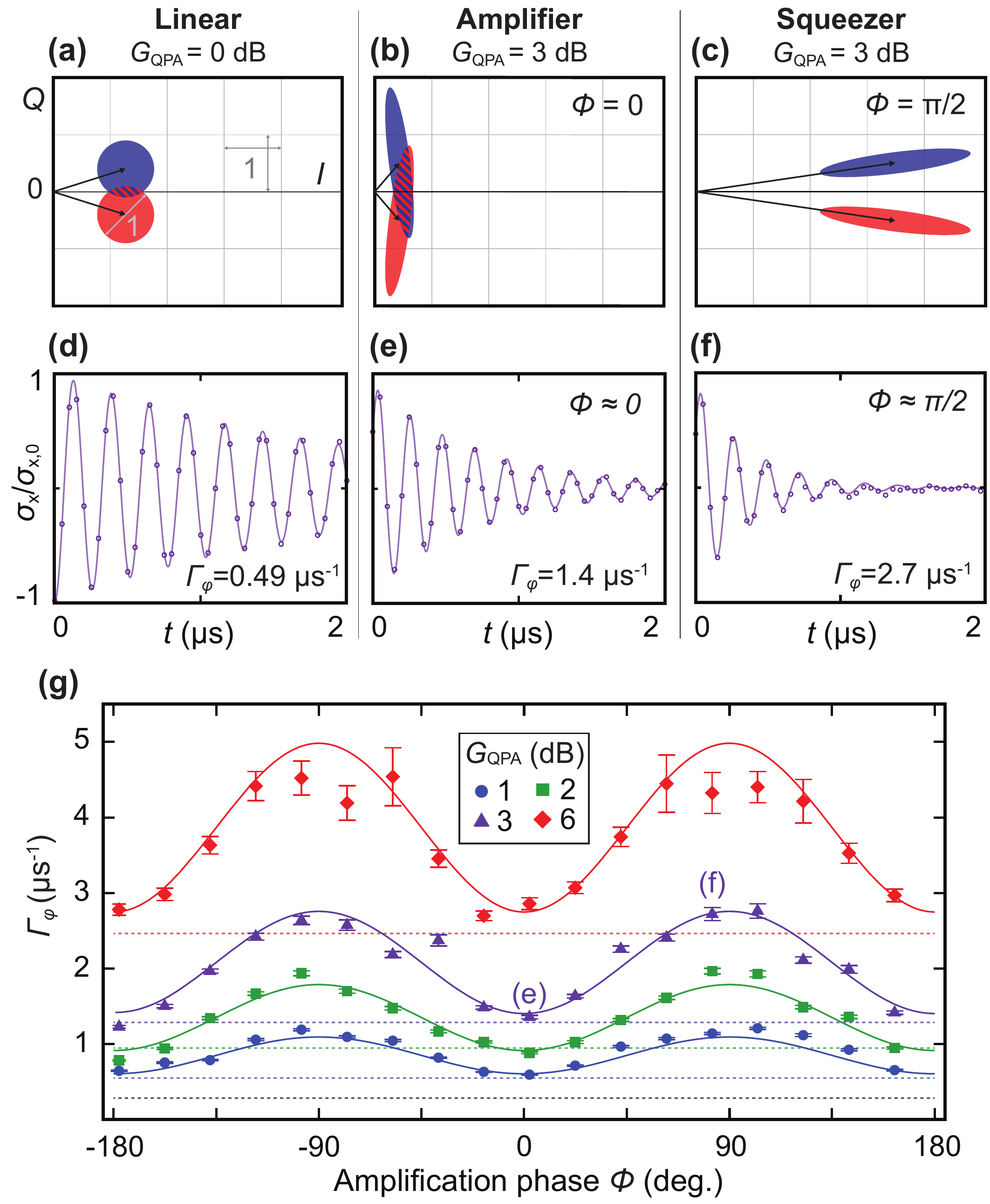}\\
  \caption{\label{dephaseFig} (a-c) Approximate QPA output-fields, calculated as in Fig. \ref{parasiteFig}. (a) With no flux pump applied, a measurement tone reflected from the QPA acquires a $\sigmaZ$ dependent phase rotation. (b) Applying a flux-pump with phase $\Phi=0$ amplifies the noise in the signal quadrature $Q$, reducing the output SNR but increasing robustness to off-chip losses. (c) The orthogonal choice $\Phi=\pi/2$ produces the opposite effects. The mean signal size is amplified slightly for either choice of $\Phi$. (d-f) Observed Ramsey traces indicate the dephasing rate $\GammaPhi$ for the drive conditions used to calculate (a-c). (g) Points indicate inferred $\GammaPhi$ values for fixed measurement power (-142 dBm) over a range of pump settings. Colored dashed lines indicate $\GammaPhiPara(\GQPA)$; the black dashed line indicates $1/T_2^*$. Solid curves are a single fit to Eq. \ref{eq:dephaseEqn}, where the only free parameter is a global phase. Dephasing is slowest (fastest) in amplifier (squeezer) mode, where the SNR at the QPA output is lowest (greatest). We attribute the discrepancy at high gain and $\Phi\approx\pm \pi/2$ to higher-order nonlinearities driven by the large photon number at this condition.}
\end{figure}

A second set of Ramsey experiments illuminates how varying on-chip gain modifies backaction during an applied weak measurement. Our theory analysis (Appendix \ref{theory}, \cite{LevitanThesis}) predicts the total dephasing to vary according to
\begin{equation} \label{eq:dephaseEqn}
\Gamma_{\phi} = \frac{2\chi^2\kappa^2P_{\text{in}}}{\hbar\omegaQPA}\left(\frac{\cos^2\Phi}{|D(-\lambda)|^2}+\frac{\sin^2\Phi}{|D(\lambda)|^2}\right) + \Gamma_{\phi, \text{parasitic}},
\end{equation}
where $P_{\text{in}}$ is the power of the measurement tone incident to the QPA, and $\GammaPhiPara$ is, notably, still given by Eq. \ref{eq:parasiticEq}. Absent any on-chip gain ($\lambda = 0$), Eq. \ref{eq:dephaseEqn} can be approximated by the more standard expression $\GammaPhi = 8\chi^2\bar{n}/\kappa + O(\frac{\chi}{\kappa})^4$, describing dephasing induced as $\sigmaZ$ information is encoded in the phase of the QPA output field (\FigThree(a), theory). Experimentally, with $\GQPA = 0$ dB we observe dephasing at rate $\Gamma_{\phi} = 0.49\ \mu \text{s}^{-1}$ (\FigThree(d)), from which we infer $P_\text{in} = -142$ dBm for this choice of drive. Keeping $P_\text{in}$ fixed, we apply a flux-pump such that $G_\text{QPA} = 3$ dB. Fig. \ref{dephaseFig}(b,c) show the expected output fields when the on-chip gain is aligned with ($\Phi=0$) or orthogonal to ($\Phi=\pi/2$) the signal quadrature. The signal size ($\langle Q_e \rangle - \langle Q_g \rangle$) is nearly constant in all three cases: since the input measurement drive lies along $I$ while the output signal lies along $Q$, the net effects of amplification and deamplification approximately cancel. In contrast, the noise fluctuations do get amplified (squeezed), such that the SNR at the QPA output depends on $\Phi$. Since amplifying the signal quadrature squeezes the photon number fluctuations in the conjugate quadrature which cause dephasing, we expect amplifier mode ($\Phi=0$) to minimize $\GammaPhi$ and squeezer mode ($\Phi=\pi/2$) to maximize it, in agreement with the comparison of Figs. \ref{dephaseFig}(e,f). Results of additional Ramsey measurements shown in Figure \ref{dephaseFig}(g) reveal the full dependence of $\GammaPhi$ on $\Phi$ and $\GQPA$, and verify the predictions of our theory model (Eq. \ref{eq:dephaseEqn}). We note that squeezer mode is also of interest as a means of improving SNR by reducing the quantum fluctuations of the output measurement field \cite{LevitanThesis,Peano2015IntracavityDeamplification,Govia2017EnhancedSuppression}; similar in-situ squeezing generation has been demonstrated in a recent optical experiment \cite{Korobko2017BeatingGeneration}.

We henceforth focus exclusively on amplifier mode ($\Phi=0$). The primary benefit of amplifier mode is that the noise floor of the QPA output signal quadrature is increased with minimal information loss, which enables greater overall efficiency by making the SNR insensitive to noise added downstream. A secondary effect is the deamplification of the mean field without deamplification of the signal; an interesting question is whether this effect, perhaps combined with injected orthogonally-squeezed vacuum, might enable a greater dispersive signal size for fixed mean intra-resonator photon number.

\section{Measurement efficiency}\label{sectMeasEff}
The total measurement efficiency is the product of on-chip efficiency and the efficiency of the rest of the measurement chain: $\etaMeas = \etaQPA \etaR$. Increasing on-chip gain, $\GQPA$, increases $\etaR$ as the amplified signal quadrature becomes robust to losses, but decreases $\etaQPA$ due to the parasitic measurement discussed in Section \ref{sectDephasing}, such that there is an optimal $\GQPA$ value maximizing $\etaMeas$ for a given measurement drive. We can write the efficiency as the ratio of empirical quantities, $\etaMeas = \GammaMeas/2\GammaPhi$. Here $\GammaMeas = \frac{\text{d}}{\text{d}t}\SNR^2/4$ is the rate at which the square of the room-temperature voltage SNR increases with integration time, or equivalently the rate at which $\sigmaZ$ information is acquired by our digitizer.

\begin{figure}
  \includegraphics[width=\columnwidth]{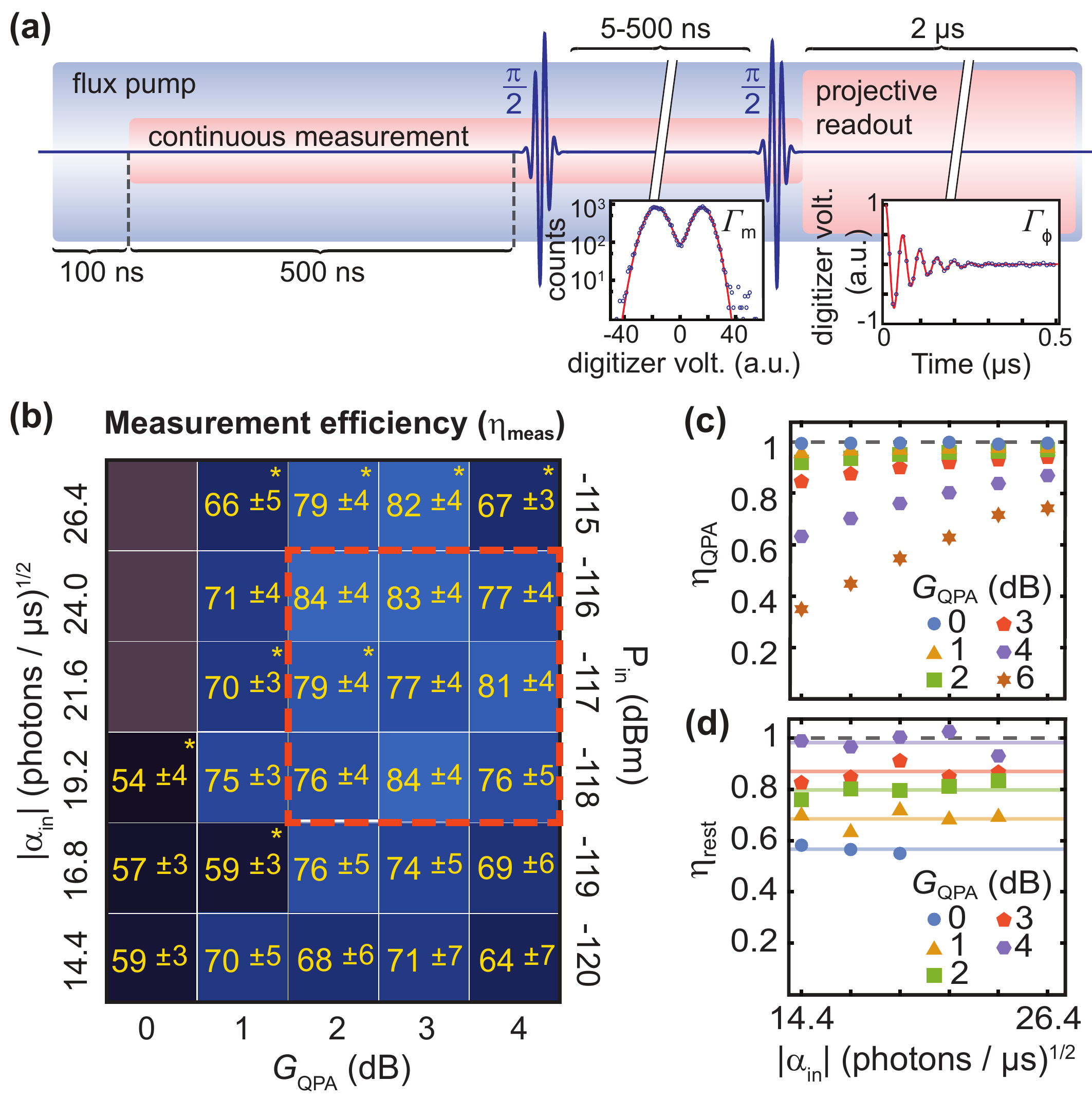}\\
  \caption{\label{etaFig} (a) Ramsey sequence used to determine $\etaMeas$. Continuous measurement of the qubit causes dephasing at rate $\Gamma_{\phi}$ (right inset). Sampling the output measurement field during the Ramsey evolution yields histograms (left inset) from which we determine SNR($t$) and thus $\GammaMeas$. (b) Inferred $\etaMeas$ values as a function of $\GQPA$ and continuous measurement amplitude, $\alphaIN$. Settings marked with an asterisk produced a spurious third histogram peak containing $> 1\%$ of the total events, which we attribute to large intra-QPA fields inducing transitions to the third transmon level. The dashed orange box surrounds the optimal operating region averaging $\etaMeas = 80\%$. (c) The on-chip efficiency $\etaQPA$, inferred by comparing $\GammaPhi$ and $\GammaPhiPara$, decreases with $\GQPA$ and increases with $\alphaIN$. (d) The off-chip efficiency, inferred by dividing $\eta$ (b) by $\etaQPA$ (c), increases with $\GQPA$ and remains roughly constant for the shown $\alphaIN$ values. Horizontal lines mark mean values to guide the eye.
}
\end{figure}

In order to determine steady-state $\etaMeas$ as a function of $\GQPA$ and measurement drive $\alphaIN = \sqrt{P_\text{in}/\hbar\omegaQPA}$, we extracted both $\GammaMeas$ and $\GammaPhi$ using the pulse sequence shown in \FigFour(a). The flux-pump was switched on in advance such that the mean intra-QPA field was deamplified along $I$ at all times, reducing the intra-QPA circulating power produced by a strong measurement drive and thus helping to minimize undesired nonlinear processes in the QPA. Next, a continuous measurement tone was turned on and sufficient time allowed to pass to ensure the cavity had stabilized before a $\pi/2$ pulse was applied to initiate Ramsey evolution. After the second $\pi/2$, $\alphaIN$ was increased to perform projective readout. The ensemble-averaged readout results for variable Ramsey evolution time were used to determine the dephasing rate $\GammaPhi$, as in the right inset of \FigFour(a). Integrating the steady-state weak-measurement record from the longest Ramsey evolution for a variable amount of time $t_\text{int} \leq 280$ ns and fitting Gaussians to the resultant histograms, we determined SNR($t_\text{int}$) and thus $\GammaMeas$. A modified Gaussian model adapted from Section III-A of \cite{Gambetta2007ProtocolsMeasurement} was used to account for $T_1$ decay  with $T_1$ fixed at the independently measured value above. Note this treatment implicitly defines $\etaMeas$ to be independent of relaxation events, such that a greater $T_1$ would result in higher readout fidelity but the same $\etaMeas$.

Sweeping measurement strength and $\GQPA$, we found an ideal operating regime, indicated by the orange dashed box in \FigFour(b), with an average $\etaMeas = 80\%$. To the left of the box, $\GQPA$ is too low to mitigate the effect of loss in circulators and other off-chip components. The bottom edge of the box is defined by the decrease in $\etaMeas$ associated with $\GammaPhiPara$ becoming a larger fraction of $\GammaPhi$ as $\alphaIN$ is decreased. The other two sides of the box are marked by the onset of non-ideal behavior evidenced by a third peak appearing in the measurement histograms. This spurious peak is not fully understood, but seems to involve population of the third transmon level driven by large intra-QPA photon numbers occurring at too low or too high $\GQPA$ (corresponding to a large mean field or a large field variance, respectively) or at too high $\alphaIN$.

\setlength{\parskip}{2pt}

With the assumption that $\GammaPhiPara$ remains independent of $\alphaIN$ at this operating point, we can express the on-chip efficiency $\etaQPA$ in terms of empirical dephasing rates as $\etaQPA = 1 - \Gamma_{\phi,\text{parasitic}}/\Gamma_{\phi},$
from which we calculate the values shown in \FigFour(c). In the absence of on-chip gain, $\etaQPA$ approaches unity as $\Gamma_{\phi,\text{parasitic}} = 1/T_2^* \ll \Gamma_{\phi}$. As gain is increased, $\Gamma_{\phi,\text{parasitic}}$ increases, resulting in lower $\etaQPA$; as the measurement strength is increased, the parasitic dephasing becomes less significant, increasing $\etaQPA$. Calculating further, we can divide these values by the $\etaMeas$ values in \FigFour(b) to estimate $\etaR$, shown in \FigFour(d). This plot is restricted to lower-power operating conditions in which the device was better behaved; over this domain, we infer that information loss downstream of the QPA decreases with $\GQPA$ and is approximately independent of $\alphaIN$, supporting our previous assumption that $\GammaPhiPara$ is likewise independent of $\alphaIN$. It is encouraging that the calculated values of $\etaR$ approach 1 at $\GQPA = 4$ dB, though the current device did not permit increasing $\alphaIN$ sufficiently to maximally benefit from this much gain. We expect this dynamic range ceiling, and thus $\etaMeas$, may be raised by reducing $\chi/\kappa$ and/or increasing the number of JPA SQUIDs \cite{Eichler2014ControllingAmplifier,Boutin2017EffectAmplifiers}, suppressing deleterious Kerr effects not included in our model.

\section{Conclusion}
We have characterized the measurement backaction on a qubit dispersively coupled to a parametric amplifier flux-pumped for gain, demonstrated how on-chip gain can mitigate off-chip sources of information loss, and observed steady-state efficiency $\etaMeas$ up to $80\%$. Going forward, incremental improvements in $\etaMeas$ may be achieved by further weakening device nonlinearities as discussed above to permit a larger measurement drive $\alphaIN$ and thus greater $\etaQPA$. A more dramatic improvement might be realized by probing the device stroboscopically \cite{Hacohen-Gourgy2016QuantumObservables}. In a linear readout resonator, stroboscopic measurement has been shown to eliminate the undesired squeezing rotations caused by dispersive coupling \cite{Eddins2017StroboscopicIllumination}; realizing the analogous effect in the QPA would close the parasitic dephasing channel such that increasing $\GQPA$ would boost $\eta_\text{rest}$ without degrading $\etaQPA$, even for small $\alphaIN$. Other potential near-term experiments include exploration of effects of on-chip gain on initial transients when switching on a measurement, and an investigation of whether combining amplifier mode with injected orthogonally-squeezed vacuum enables greater dispersive readout SNR for a fixed intra-resonator photon number. Longer term, we envision the QPA to be an enabling technology for applications demanding signal-to-noise ratios approaching the quantum limit, such as measurement-based quantum feedback \cite{Li2013OptimalityImperfections,Martin2015DeterministicFeedback,Martin2017WhatFeedback} or further studies of individual quantum trajectories, perhaps with extension to multi-qubit experiments via a chip layout similar to \cite{Schmitt2014MultiplexedAmplifiers} followed by a broadband off-chip amplifier.

\begin{acknowledgments}
The authors thank R. Vijay, S. Hacohen-Gourgy, E. Flurin, and L. Martin for useful discussions, and thank MIT Lincoln Labs for fabrication of the JTWPA with support from the LPS and IARPA. Work was supported by the Army Research Office under Grant No. W911NF-14-1-0078. A.A.C. and L.C.G.G. acknowledge support from the the AFOSR MURI FA9550-15-1-0029. A.E. acknowledges support from the Department of Defense through the NDSEG fellowship program.
\end{acknowledgments}

\appendix

\section{Experimental Methods}

\subsection{Device fabrication}
The device used for this paper is patterned on $>8000$ $\Omega$-cm intrinsic Si using photolithography and subsequent plasma etching of 100 nm thick e-beam evaporated Al.  Al/AlO$_x$/Al junctions for the transmon and SQUIDs are defined in separate steps with e-beam lithography and subsequent double-angle evaporation. Adhering to the constraints of the pre-defined bond pads and QPA interdigitated capacitor, we were able to reduce the radiative loss of the qubit by introducing electrically-floating metal shielding around the qubit capacitor paddles (white in \FigOne(c)). The chip is mounted on OFHC copper enclosed in an Al package, surrounded by cryoperm, and mounted to the base stage of a dilution refrigerator at $\leq$35 mK. A copper wire fed through a high aspect-ratio hole in the aluminum helps thermalize the interior OFHC mount to the base stage. Flux to tune the QPA was applied with an off-chip coil.

\subsection{Circuit parameters}\label{parametersAppendix}

Circuit parameters changed slightly after thermal cycling, and also as the QPA was flux-biased to different operating frequencies $\omegaQPA$. The table lists precise parameter information corresponding to the three data figures of the main text. The change in $\chi$ between cooldowns is not well understood.

\begin{center}
    \begin{tabular}{| l | l | l | l |}
    \hline
    \  & Fig. 2 & Fig. 3 & Fig. 4 \\ \hline
    Cooldown & B & A & B \\ \hline
    $\omegaQPA/2\pi$ (GHz) & 6.740 & 6.740 & 6.700 \\ \hline
    $\kappa/2\pi$ (MHz) & 25.4 & 25.7 & 28.6 \\ \hline
    $\omegaQ/2\pi$ (GHz) & 4.271 & 4.274 & 4.271 \\ \hline
    $\chi/2\pi$ (MHz) & 1.9 & 1.7 & 2.0 \\ \hline
    \end{tabular}
\end{center}

\subsection{Detailed wiring diagram}
\begin{figure}[t]
  \includegraphics[width=\columnwidth]{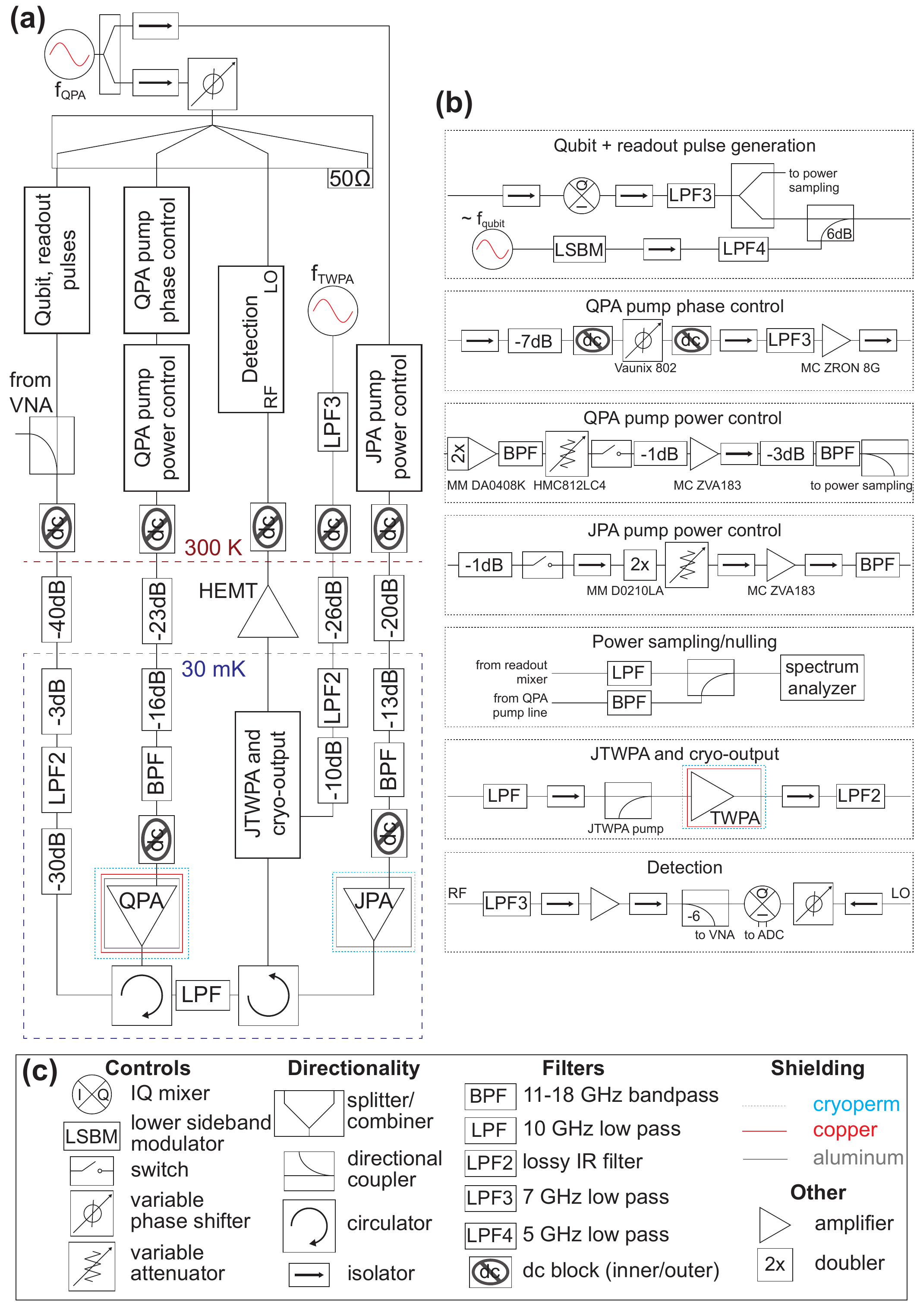}\\
  \caption{\label{fullSetup} Detailed experimental wiring diagram. 
}
\end{figure}

Figure \ref{fullSetup} shows a complete diagram of the experiment. An experimental overview appears in (a), with individual subsystems detailed in (b) and a component legend given in (c). Most centrally, a microwave generator at $\omegaQPA$ (estimated by measuring the QPA resonance frequency while Rabi-driving the qubit) is split to drive four phase-sensitive processes: dispersive qubit measurement, on-chip amplification in the QPA, off-chip amplification in the JPA, and room-temperature demodulation of the output measurement signal. The amplification processes are highly sensitive to changes in applied pump power, such that the small change in insertion-loss associated with adjusting the phase of a phase-shifter would problematically change the amplifier gain. Several technical solutions are possible; for the QPA, we use a spectrum analyzer to ensure power-flatness as we programmatically step the phase of the flux-pump. For the JPA, we change the effective amplification phase by phase-shifting all other tones at $\omegaQPA$ while leaving the JPA pump unchanged. To realize amplifier-mode operation of the QPA, the phase of the QPA pump was first chosen to minimize $\GammaPhi$, and then the phase of the off-chip JPA pump was adjusted to maximize SNR. Stark shifts were measured in advance for all $\GQPA, \alphaIN$ settings used to produce the data in \FigFour, and qubit pulse frequencies and amplitudes were programmatically adjusted accordingly at each setting. Small superconducting coils (not shown) were used to apply dc-flux biases to the QPA and JPA to tune their resonance frequencies. A vector network analyzer was used to characterize device resonance frequencies and gains. Cross-talk effects of the JPA flux-pump on the JTWPA became apparent at high JTWPA gain despite the intermediary low-pass filter, degrading JTWPA performance. These effects were suppressed by operating the JTWPA at reduced gain ($\sim 15$ dB).

\begin{figure*}[t]
  \includegraphics[width=\textwidth]{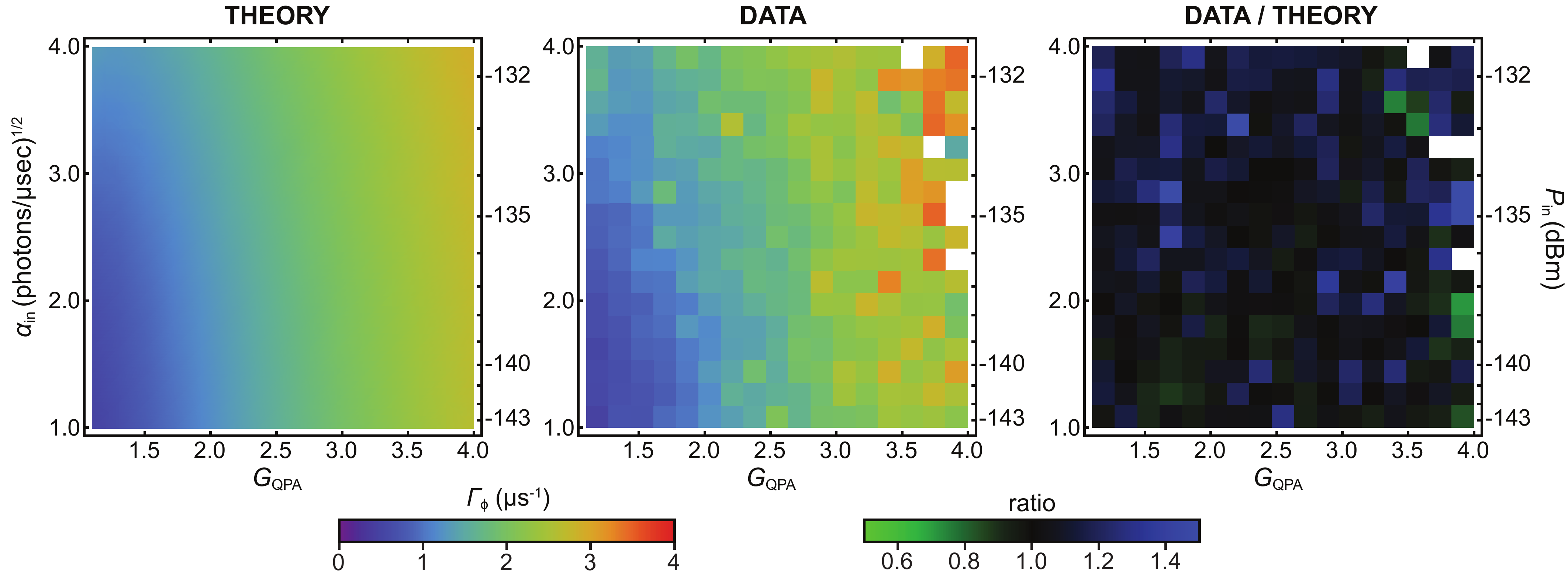}\\
  \caption{\label{dephasing2d} Dephasing rate $\GammaPhi$ as a function of measurement strength $\alphaIN$ and on-chip gain $\GQPA$. The left panel shows the prediction of Eq. \ref{eq:dephaseEqn}, while the central panel shows the $\GammaPhi$ values inferred from observation. The ratio of the data to the theory prediction is displayed in the right panel.
}
\end{figure*}

\subsection{$T_1$ vs on-chip gain}
\begin{figure} [h!]
\includegraphics[width=0.9\columnwidth]{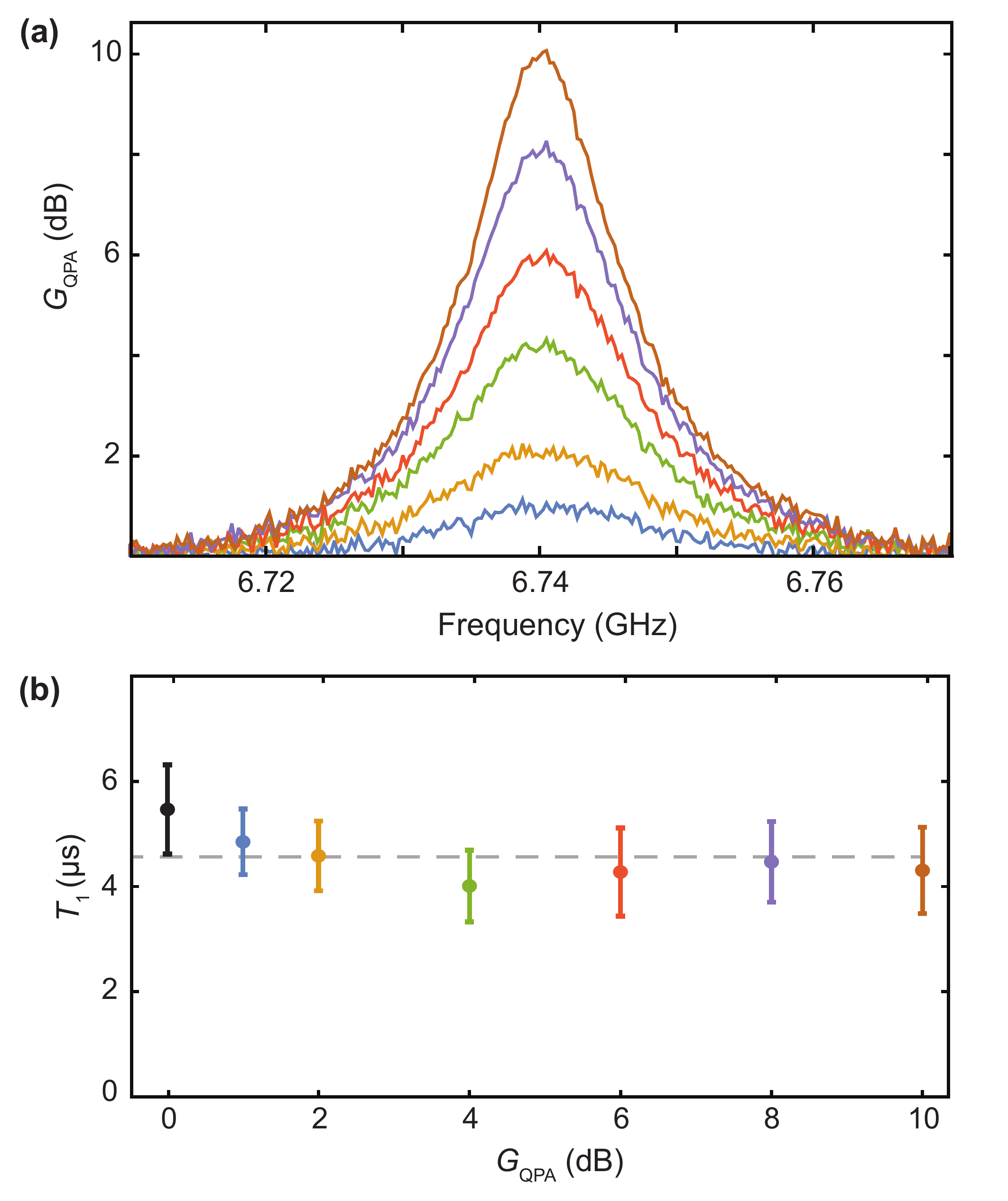}\\
\caption{\label{T1vsGain} (a) Gain profiles of the QPA for varying amounts of on-chip gain $\GQPA$ as measured with a vector network analyzer. (b) Qubit $T_1$ measured at several values of $\GQPA$ corresponding to the color-coded gain profiles in (a). Measurements were repeated for approximately $\sim 12$ hours, during which time the order in which the gain settings were cycled through was repeatedly randomized. Error bars indicate the standard deviations of all results at each $\GQPA$ setting.}
\end{figure}

We briefly investigated the effect of on-chip gain on the qubit lifetime $T_1$. Results are shown in Fig. \ref{T1vsGain}. The data suggest a small decrease in $T_1$ when gain is turned on, with no clear dependence as gain is further increased.

\subsection{Amplifier-mode dephasing vs $P_\text{in}$}
Extending the measurement-backaction data presented in \FigThree, we fixed the QPA pump phase to operate the QPA in amplifier mode ($\Phi=0$) and recorded $\GammaPhi$ for variable on-chip gain and measurement strength $\alphaIN$. The results are displayed alongside the theory prediction of Eq. \ref{eq:dephaseEqn} in Fig. \ref{dephasing2d}. Good agreement is seen for low measurement strength and on-chip gain. At high drive strengths or high gains, excess dephasing is observed intermittently, i.e. for some experimental executions. At intermediate drive strength, near the center of the plot, this undesired behavior appears to be reduced as $\GQPA$ increases from $\sim$1 dB to $\sim2-3$ dB.

\section{Theoretical derivations} \label{theory}

\subsection{Dephasing with on-chip gain}

Our goal is to derive an analytic expression for the qubit dephasing rate in the long-time limit, without assuming that the dispersive coupling $\chi$ is weak. We start with the master equation
\begin{align}
  \nonumber\dot{\hat{\rho}} = &-i \left[ \hat{H}, \hat{\rho} \right] + \kappa \mathcal{D} \left[ \hat{a} \right] \hat{\rho} \\ &+ \frac{1}{T_1} \mathcal{D} \left[ \hat{\sigma}_- \right] \hat{\rho} + \frac{1}{2T_2} \mathcal{D} \left[ \hat{\sigma}_z \right] \hat{\rho}, \label{eqn:ME}
\end{align}
where $\kappa$ is the resonator decay rate, $T_1$ and $T_2$ are the relaxation and pure dephasing times of the qubit, $\hat{\sigma}_-$ is the qubit lowering operator, and $\mathcal{D} \left[\hat{O}\right] \hat{\rho} = \hat{O} \hat{\rho} \hat{O}^{\dagger} - \frac{1}{2} \left\{\hat{O}^{\dagger} \hat{O}, \hat{\rho} \right\}$ is the usual dissipator. In a frame where the qubit rotates at its bare frequency $\omega_{\rm q}$, and the resonator at its static flux-biased frequency $\omega_{\rm QPA}$, the Hamiltonian is
\begin{align}
  \hat{H} = \frac{i \lambda}{2} \left( \hat{a}^{\dagger 2} - \hat{a}^2\right) + \chi \hat{a}^{\dagger} \hat{a} \hat{\sigma}_z
		+ \sqrt{\kappa} \left( \alpha_{\rm in} \hat{a}^{\dagger} + \alpha_{\rm in}^{*} \hat{a} \right),
\end{align}
which contains the QPA dynamics of $\hat{H}_{\rm QPA}$ from Eq.~\ref{eq:QPAHam}, and the coherent measurement drive on the QPA resonator, characterized by the drive amplitude $\alpha_{\rm in}$.

The dephasing rate quantifies the decay of the qubit coherence in the long time limit, described by the decay of the qubit off-diagonal matrix elements of the full density matrix. If we write the full density matrix as

\begin{align}
  \hat{\rho} = \sum_{\mu,\nu \in\{\uparrow,\downarrow\}}\hat{\rho}_{\mu\nu}\otimes\ket{\mu}\bra{\nu}
\end{align}
where $\hat{\rho}_{\mu\nu}$ is an operator on the resonator Hilbert space and $\ket{\downarrow}$ and $\ket{\uparrow}$ are the ground and excited states of the qubit,  then the qubit dephasing rate is fully captured by the evolution of the part of the density matrix proportional to $\ket{\uparrow}\bra{\downarrow}$ (or its Hermitian conjugate). Thus, we are interested in the evolution of the operator $\hat{\rho}_{\uparrow\downarrow}$. As is standard, we define the dephasing rate as
\begin{align}
  \Gamma_{\phi} = \lim_{t \rightarrow \infty} \frac{-{\rm ln}\left({\rm Tr}\left[\hat{\rho}_{\uparrow\downarrow}(t)\right]\right)}{t}, \label{eqn:Drate}
\end{align}
which captures the exponential decay of the qubit coherence in the long-time limit.

From Eq.~\ref{eqn:ME} we calculate the evolution equation for the operator $\hat{\rho}_{\uparrow\downarrow}$
\begin{align}
  \nonumber \dot{\hat{\rho}}_{\uparrow\downarrow} &= \left[ \frac{\lambda}{2} (\hat{a}^{\dagger} \hat{a}^{\dagger} - \hat{a} \hat{a}) - \sqrt{\kappa} (\alpha_{\rm in} \hat{a}^{\dagger} - \alpha_{\rm in}^{*} \hat{a}), \hat{\rho}_{\uparrow\downarrow} \right]\\
  &- i \chi \left\lbrace \hat{a}^{\dagger} \hat{a}, \hat{\rho}_{\uparrow\downarrow} \right\rbrace
		+ \kappa \mathcal{D} [\hat{a}] \hat{\rho}_{\uparrow\downarrow} - \left(\frac{1}{2T_1} + \frac{1}{T_2}\right)\hat{\rho}_{\uparrow\downarrow}. \label{eqn:UpDown}
\end{align}
Note that this equation is not trace preserving, as it does not describe evolution of a valid density matrix. Extending beyond the results of Ref.~\cite{LevitanThesis}, we have included the effect of qubit relaxation $(T_1)$ in the evolution of $\hat{\rho}_{\uparrow\downarrow}$, and while this means the evolution of the qubit is no longer QND, the resulting equation for $\hat{\rho}_{\uparrow\downarrow}$ remains closed on itself, and can be solved analytically.

The first step in solving Eq.~\ref{eqn:UpDown} is to remove the exponential decay caused by the qubit incoherent dynamics, and we do so by defining $\hat{\rho}'_{\uparrow\downarrow} = e^{t/T_2^*}\hat{\rho}_{\uparrow\downarrow}$, where $T_2^* = 2T_1T_2/(2T_1 + T_2)$ introduced in the main text describes the intrinsic dephasing of the qubit. The evolution equation for $\hat{\rho}'_{\uparrow\downarrow}$ describes the qubit dephasing due to interaction with the resonator, and has the same form as Eq.~\ref{eqn:UpDown}, but without the last term (proportional to $1/T_2^*$).

To solve the evolution equation for $\hat{\rho}'_{\uparrow\downarrow}$, it is more convenient to move to the Wigner representation, and obtain a partial differential equation for $W_{\rm \uparrow\downarrow} (x, p; t)$, the Wigner function representation of $\hat{\rho}'_{\uparrow\downarrow}$ \cite{WahyuUtami2008EntanglementSystem}. As Eq.~\ref{eqn:UpDown} contains terms at most quadratic in $\hat{a}$ and $\hat{a}^\dagger$ it is possible to solve this PDE with a Gaussian ansatz.

The Gaussian ansatz reduces Eq.~\ref{eqn:UpDown} to a set of coupled ODE's for the means, variances and overall norm of $\hat{\rho}'_{\uparrow\downarrow}$. After solving these in steady state (see Ref.~\cite{LevitanThesis} for further details), with the coherent drive defined by
\begin{align}
  \alpha_{\rm in} = \sqrt{\frac{P_{\rm in}}{\hbar\omega_{\rm QPA}}}\left(\cos\left(\Phi\right) + i\sin\left(\Phi\right)\right), \label{eqn:alpha}
\end{align}
we can then use Eq.~\ref{eqn:Drate} to define the dephasing rate, which gives the expression found in Eq.~\ref{eq:dephaseEqn}. (By defining $\alphaIN$ in terms of $\Phi$ here, we implicitly fix the phase of the pump, in contrast to the convention used in the main text).

\subsection{Measurement rate with on-chip gain}

We now briefly outline the theoretical calculations of the measurement rate, and for further details the interested reader should consult chapter 3 of Ref.~\cite{LevitanThesis}. From standard input-output theory, the Heisenberg-Langevin equation for the resonator operator $\hat{a}$ in a frame rotating at the bare resonator frequency $\omega_{\rm QPA}$ is
\begin{align}
  \dot{\hat{a}} = \left( -i \chi \hat{\sigma}_z - \frac{\kappa}{2} \right) \hat{a}   + \lambda \hat{a}^{\dagger} - \sqrt{\kappa} \hat{a}_{\rm in}, \label{eqn:HL}
\end{align}
where $\hat{a}_{\rm in}$ is the input field to the resonator. In our case, for dispersive measurement of the qubit this is ideally a coherent state, such that $\left<\hat{a}_{\rm in}\right> = \alpha_{\rm in}$, with $\alpha_{\rm in}$ defined in Eq.~\ref{eqn:alpha}. For the purposes of the intra-resonator dynamics and calculation of the measurement rate we can treat the qubit operator $\hat{\sigma}_z$ as a classical real variable $\sigma = \pm 1$, corresponding to the ground or excited state of the qubit in the $\hat{\sigma}_z$ basis. Doing so allows us to solve Eq.~\ref{eqn:HL} exactly, and from this solution extract the measurement rate.

We consider two modes of operation for the QPA, ``amplifier mode'', where the input field aligns with the direction of squeezing ($\Phi = 0$), and ``squeezer mode'', where the input field aligns with the direction of amplification ($\Phi = \pi/2$). In amplifier mode, we use the QPA to amplify the size of the signal created by the qubit, which comes at the cost of also amplifying the noise in the cavity output field. In squeezer mode, we use the QPA to squeeze the noise in the quadrature containing qubit information, and while the noise can be heavily squeezed, the signal produced by the qubit is only squeezed at most by a factor of two. For a fixed input photon flux, the steady-state intra-resonator photon number is not the same for both modes of operation, but is independent of the qubit state in both cases.

The output field is related to the input field by the standard input-output relation $\hat{a}_{\rm out} = \hat{a}_{\rm in} + \sqrt{\kappa}\hat{a}$, and from the output mode we define the measured signal operator by
\begin{align}
  \hat{Q}(t) = \frac{e^{-i \delta} \hat{a}_{\rm out} + e^{i \delta} \hat{a}_{\rm out}^{\dagger}}{\sqrt{2}}, \label{eqn:I}
\end{align}
where the angle $\delta$ parameterizes the quadrature measured. We must choose the measured quadrature such that it is out-of-phase with the input coherent signal (as the qubit information will be contained in the out-of-phase quadrature), such that $\abs{\delta - \Phi} = \pi/2$.

As we are interested in the long-time limit of the QPA dynamics, rather than the SNR we will calculate the measurement rate, defined by
\begin{align}
  \Gamma_{\rm meas} &\equiv \lim_{\tau \rightarrow \infty} \frac{{\rm SNR}^2 (\tau)}{2 \tau} = \frac{1}{4} \frac{\left(\left<\hat{Q}\right>_{\uparrow} - \left<\hat{Q}\right>_{\downarrow}\right)^2}{(\bar{S}_{QQ, \uparrow} [0] + \bar{S}_{QQ, \downarrow} [0])} \label{eqn:GammaM},
\end{align}
where $\left<.\right>_{\nu}$ indicates that the expectation value is taken with respect to the cavity in steady-state and the qubit in state $\ket{\nu}$ for $\nu\in \{\uparrow,\downarrow\}$, corresponding to $\sigma = \pm 1$ respectively. $\bar{S}_{QQ, \nu} [\omega]$ is the symmetrized noise power of the detected quadrature at frequency $\omega$, defined in the standard way \cite{LevitanThesis}.

The measurement rate will depend on what mode the QPA is operated in (i.e.~the angle $\Phi$), and for our two operation modes the measurement rates are
\begin{align}
  &\Gamma_{\rm meas}^{\rm amp}= \frac{\frac{\chi^2 \kappa |\alpha|^2}{(\frac{\kappa}{2} - \lambda)^2 + \chi^2}}
			{\frac{1}{2}\frac{ \left[ (\frac{\kappa}{2} + \lambda)^2 - \chi^2 \right]^2 + \chi^2 \kappa^2}{\left(\frac{\kappa^2}{4} - \lambda^2 + \chi^2 \right)^2} + \bar{n}_{\rm add}},\label{eqn:GammaA} \\
  &\Gamma_{\rm meas}^{\rm sqz}= \frac{\frac{\chi^2 \kappa |\alpha|^2}{(\frac{\kappa}{2} + \lambda)^2 + \chi^2}}
          {\frac{1}{2}\frac{ \left[ (\frac{\kappa}{2} - \lambda)^2 - \chi^2 \right]^2 + \chi^2 \kappa^2}{\left(\frac{\kappa^2}{4} - \lambda^2 + \chi^2 \right)^2} + \bar{n}_{\rm add}},\label{eqn:GammaS}
\end{align}
where we have added by hand a noise term $\bar{n}_{\rm add}$ to quantify noise added to the signal downstream of the QPA. For a fair comparison we parametrize the rates in terms of a constant intra-resonator photon number $\abs{\alpha}^2$, which we note requires different input photon flux for the two operation modes.

From Eqs.~\ref{eqn:GammaA} and \ref{eqn:GammaS} we see that in both modes of operation the output contains amplified noise, no matter the value of $\lambda$. While this is by design in amplifier mode, in squeezer mode it is na\"ively unexpected, and is a result of interaction with the qubit. The dispersive interaction results in a qubit-dependent phase shift on the field exiting the resonator, such that it no longer perfectly interferes with the promptly reflected field. The effect of this is a mixing of the squeezed and amplified noise, such that all quadratures contain noise contributions from both.

However, for very small $\chi/\kappa$, squeezer mode operation does not suffer from this unwanted mixed-in amplified noise, as can be seen when we write the measurement rates to leading order in $\chi/\kappa$
\begin{align}
  &\Gamma_{\rm meas}^{\rm amp} \approx \frac{2\chi^2|\alpha|^2(1+\sqrt{G_0})^2}{\kappa(G_0+2\bar{n}_{\rm add})}, \label{eqn:GammaA0} \\
  &\Gamma_{\rm meas}^{\rm sqz} \approx \frac{2\chi^2|\alpha|^2(1+1/\sqrt{G_0})^2}{\kappa(1/G_0+2\bar{n}_{\rm add})}, \label{eqn:GammaS0}
\end{align}
where we have defined $\sqrt{G_0} = (\kappa/2 + \lambda)/(\kappa/2 - \lambda)$, with $G_0 = 1$ for zero gain. Both measurement rates should be contrasted with the zero gain measurement rate
\begin{align}
  \nonumber\Gamma_{\rm meas}^{0} &= \frac{2\chi^2 \kappa |\alpha|^2}{\left(\frac{\kappa^2}{4} + \chi^2\right)\left[1 + 2\bar{n}_{\rm add}\right]} \\ &\xrightarrow{\chi/\kappa \ll 1} \frac{8\chi^2 |\alpha|^2}{\kappa(1+2\bar{n}_{\rm add})}, \label{eqn:GammaSt}
\end{align}
found for a standard linear-resonator setup, or when the QPA is operated with zero gain.

In the ideal limit, where $\bar{n}_{\rm add} = 0$, amplifier mode offers little to no advantage over zero gain, as both the signal and noise are amplified by the same factor at large gain, which can be seen by comparing Eq.~\ref{eqn:GammaA0} for $\bar{n}_{\rm add} = 0$ to Eq.~\ref{eqn:GammaSt}. However, in this case squeezer mode can be advantageous, as in the large gain limit the noise is drastically reduced, while the signal is relatively unaffected. In particular, for $\chi/\kappa \ll 1$, comparing Eq.~\ref{eqn:GammaS0} for $\bar{n}_{\rm add} = 0$ to Eq.~\ref{eqn:GammaSt}, we see that the measurement rate is enhanced by a large factor proportional to $G_0$. Accounting for effects beyond first order in $\chi/\kappa$ by using the full expression of Eq.~\ref{eqn:GammaS}, we find that the squeezer mode measurement rate is enhanced by the factor $\Gamma_{\rm meas}^{\rm sqz}/\Gamma_{\rm meas}^{0} = \kappa/\chi$ at the optimal value of $\lambda$.

Conversely, in the non-ideal situation where $\bar{n}_{\rm add}$ is large, squeezer mode offers no advantage, as the noise can never be reduced below the noise floor set by $\bar{n}_{\rm add}$, as clearly indicated by Eq.~\ref{eqn:GammaS0}. In this situation amplifier mode is beneficial, as by amplifying both the signal and noise leaving the QPA, the output becomes insensitive to noise added downstream (concretely, $G_0 \gg \bar{n}_{\rm add}$ in the denominator of Eq.~\ref{eqn:GammaA0}). As shown in the main text, this is the mode of operation we find gives the greatest efficiency for our current setup.



%


\end{document}